\documentclass[prb,twocolumn,showpacs,preprintnumbers,amsmath,amssymb]{revtex4}
\usepackage{graphicx}
\usepackage{dcolumn}
\usepackage{bm}
\usepackage{stmaryrd}
\usepackage{amssymb}
\usepackage{amsfonts}
\usepackage{amsmath}

\begin{document}
\title{Effects of interaction and polarization on spin-charge separation: A time-dependent spin-density-functional theory study}
\date{\today}

\author{Gao Xianlong}
\email{gaoxl@zjnu.edu.cn}
\affiliation{Department of Physics, Zhejiang Normal University, Jinhua 321004, Zhejiang Province, China}

\begin{abstract}
We calculate the nonequilibrium dynamic evolution of a one-dimensional system of two-component fermionic atoms after a strong local quench by using a time-dependent spin-density-functional theory. The interaction quench is also considered to see its influence on the spin-charge separation. It is shown that the charge velocity is larger than the spin velocity for the system of on-site repulsive interaction (Luttinger liquid), and vise versa for the system of on-site attractive interaction (Luther-Emery liquid). We find that both the interaction quench and polarization suppress the spin-charge separation.
\end{abstract}
\maketitle

\section{Introduction}
While the nonequilibrium dynamic evolution of quantum systems has long been extensively studied,~\cite{Mattis}
progress is hindered by the tremendous difficulties in solving the nonequilibrium quantum many-body Schr\"{o}dinger equation.
This situation is going to be changed due to the progress in experiments and the development in numerical methods.

On the experimental side, the development in manipulating ultracold atomic gases makes it feasible to study strongly correlated systems with time-varying interactions and external potentials and in out-of-equilibrium situations. The high controllability in ultracold atomic-gases' systems provides an ideal testbed to observe the long-time evolution of strongly correlated quantum many-body systems, and to test theoretical predictions, such as the Bloch oscillation,~\cite{Bloch} the absence of thermalization in nearly integrable one-dimensional (1D) Bose gases,~\cite{Kinoshita} and the expansion of BEC in a random disorder after switching off the trapping potential.~\cite{Anderson} These efforts allow us to study the nonequilibrium dynamics of strongly correlated systems from a new perspective.

Numerically, many techniques have been developed, such as, the time-adaptive density-matrix renormalization group (t-DMRG),~\cite{Schollwoek} the time-dependent numerical renormalization group,~\cite{Costi} continuous-time Monte Carlo algorithm,~\cite{Eckstein} and time-evolving block decimation method.~\cite{Vidal} Time-dependent spin-density-functional theory (TDSDFT) has been proved to be a powerful numerical tool beyond the linear-response regime in studying the interplay between interaction and the time-dependent external potential.~\cite{gaoprb78Liwei,verdozzi_2007} More tests of the performance of TDSDFT will be done in this paper on the polarized system with attractive or repulsive interactions. Compared to the algorithms, such as the t-DMRG, this technique gives numerically inexpensive results for large lattice systems and long-time evolution, but with difficulties in calculating some properties, such as, the correlation functions.

The 1D bosonic or fermionic  systems accessible by the present ultracold experiments,~\cite{moritz,bosonic} are exactly solvable in some cases~\cite{Gaudin} and can be used to obtain a thorough understanding of the many-body ground-state and the dynamical properties. The nonequilibrium problems in 1D system are especially remarkable in which the 1D systems are strongly interacting, weakly dissipative, and lack of thermalization.~\cite{Sutherland} The 1D systems, belonging to the universality class described by the Luttinger-liquid theory,
have its particularity in its low-energy excitations, characterized by charged, spinless excitations and neutral, spin-carrying collective excitations. Generically, the different dynamics is determined by the velocities of the charge and spin collective excitations, which has been verified experimentally in semiconductor quantum wires by Auslaender {\it et al.}.~\cite{Auslaender} The possibility of studying these phenomena experimentally in 1D two-component cold Fermi gases,~\cite{moritz} where "spin" and "charge" refer, respectively, to the density difference and the total atomic mass density of the two internal atomic states, was first highlighted by Recati {\it et al.}~\cite{recati_prl_2003}. The different velocities for spin and charge in the propagation of wave packets have been demonstrated by Kollath et al.~\cite{kollath,Kollath2} in a numerical t-DMRG study of the 1D Fermi-Hubbard model, by Kleine {\it et al.}~\cite{kleine} in a similar study of the two-component Bose-Hubbard model, and, analytically, by Kecke {\it et al.}~\cite{kecke_prl_2005} for interacting fermions in a 1D harmonic trap.
Exact diagonalization and quantum Monte Carlo simulations are also used in studying the spin and charge susceptibilities of the Hubbard model.~\cite{Jagla,Zacher}
Dynamic structure factors of the charge density and spin are analyzed for the partially spin-polarized 1D Hubbard model with strong attractive interactions using a time-dependent density-matrix renormalization method.~\cite{Huse} The spin-charge separation is well addressed for this system.~\cite{Huse}
We would like to mention here that a genuine observation of spin-charge separation requires one to explore the single-particle excitation, which is studied recently in simulating the excitations created by adding or removing a single particle.~\cite{Kollath2,Ulbricht}

The nonequilibrium dynamics in 1D systems has attracted a growing attention in the possible equilibrium properties after an external perturbation and the changes in physical quantities after the quench.~\cite{quench,kollath,Karlsson} The dynamic phase transition and different relaxation behavior are studied with a sudden interaction quench~\cite{Eckstein,Barmettler}. The relation between the thermalization and the integrability in 1D system is well addressed.~\cite{Rigol}
The real-time evolution for the magnetization in the 1D spin chain is also studied in great details using the t-DMRG.~\cite{Langer}

In this paper, we study the 1D system under an instantaneous switching off a strong local potential or on-site interactions, namely, a sudden quantum quench is considered. The strong local potential creates Gaussian-shaped charge and/or spin accumulations at some position in space. After the quantum quench, the time-evolution of spin and charge densities is then calculated at later times. We tackle this problem using TDSDFT based on an adiabatic local spin density approximation (ALSDA).

The contents of the paper are as follows. In Sec.~\ref{sect:model}, we introduce the model: a time-dependent lattice Hamiltonian that we use to study spin-charge separation and quench dynamics. Then we briefly summarize the self-consistent lattice TDSDFT scheme that we use to deal with the time-dependent inhomogeneous system. In Sec.~\ref{sect:numerical_results}, we report and discuss our main numerical results. At last, a concluding section summarizes our results.

\section{Model and the method}
\label{sect:model}
We consider a two-component repulsive/attractive Fermi gas with
$N_f$ atoms loaded in a 1D optical lattice with $N_s$ lattice sites.
At time $t\leq 0$, a localized spin- and charge-density perturbation is created by switching on slowly the local potential,
such that the system is in the ground state of the system with the additional potential. At $t=0^+$, the
localized potential is removed abruptly and/or the on-site interaction is switched off instantaneously. This system is modeled by a time-dependent Fermi-Hubbard Hamiltonian as follows:
\begin{eqnarray}\label{eq:hubbard}
    {\hat {\cal H}}(t)&=&-\gamma\sum_{i,\sigma}({\hat c}^{\dagger}_{i\sigma}
    {\hat c}_{i+1\sigma}+{\rm H}.{\rm c}.)+U(t)\sum_i
    {\hat n}_{i\uparrow}{\hat n}_{i\downarrow}\nonumber\\
    &+&\sum_{i, \sigma} V_{i\sigma}(t) {\hat n}_{i\sigma}~.
\end{eqnarray}
Here $\gamma$ is the hopping parameter, ${\hat c}^{\dagger}_{i\sigma}$ (${\hat c}_{i\sigma}$) creates (annihilates)
a fermion in the $i$th site ($i \in[1,N_s]$), $\sigma=\uparrow,\downarrow$ is a pseudospin-$1/2$ degree-of-freedom
(hyperfine-state label), $U(t)$ is the time-dependent on-site Hubbard interaction of negative or attractive nature, and
${\hat n}_{i\sigma}={\hat c}^{\dagger}_{i\sigma}{\hat c}_{i\sigma}$. We also introduce for future purposes the local number operator ${\hat n}_{i}=\sum_\sigma {\hat n}_{i\sigma}$ and the local spin operator ${\hat s}_{i}=\sum_\sigma \sigma {\hat n}_{i\sigma}/2$.

The external time-dependent potential $V_{i\sigma}(t)=V^{\rm ext}_{i\sigma} \Theta(-t)$, which simulates the spin-selective focused laser-induced potential. $\Theta(t)$ is the Heaviside step function which relates the quench dynamics to the modification of the local potential. $\Theta(-t)=0$ for $t>0$. $V^{\rm ext}_{i\sigma}$ is taken to be of the following Gaussian form:
\begin{eqnarray}\label{eq:ext_pot}
V^{\rm ext}_{i\sigma}&=&W_\sigma\exp{\left\{-\frac{[i-(N_s+1)/2]^2}{2\alpha^2}\right\}} ~.
\end{eqnarray}
Here $W_\sigma$ is the amplitude of the local potential. We discuss the system of conserved particle number in the canonical ensemble. The number of atoms for spin up and spin down is, $N_\uparrow$ and $N_\downarrow$, respectively. The polarization is defined as $p=(N_\uparrow-N_\downarrow)/N_f$. The on-site interaction and $W_\sigma$ are scaled in units of $\gamma$ as, $u=U/\gamma$ and $w_\sigma=W_\sigma/\gamma$, respectively.

A powerful theoretical tool to investigate the dynamics of many-body systems in the presence of time-dependent
inhomogeneous external potentials, such as that in Eq. (\ref{eq:hubbard}), is TDSDFT,~\cite{Giuliani_and_Vignale,marques_2006} based on the Runge-Gross theorem~\cite{rgt} and on the time-dependent single-particle Kohn-Sham equations. The complication of the problem is hidden in the unknown time-dependent exchange and correlation (xc) potential.
Most applications of TDSDFT use the simple adiabatic local spin-density approximation for the dynamical xc potential,~\cite{Giuliani_and_Vignale, zangwill} which has often been proved to be successful in studying the real-time evolution.~\cite{marques_2006}
In this approximation, one assumes that the time-dependent xc potential is just the static xc potential evaluated at the instantaneous density, where the xc potential is local in time and space. The static xc potential is then treated within the static local spin-density approximation. Very attractive features of the ALSDA are its extreme simplicity, the ease of implementation, and the fact that it is not restricted to mean-field approximation and small deviations from the ground-state density, i.e., to the linear response regime. The dynamics induced by the strong local perturbation discussed here cannot be dealt with the theory based on the linear response while TDSDFT is a good candidate.

We here employ a lattice version of spin-density-functional theory (SDFT) and TDSDFT.~\cite{gaoprb78Liwei} In short, for times $t\leq 0$, the spin-resolved site-occupation profiles can be calculated by means of a static SDFT. For times $t>0$, we calculate the time evolution of spin-resolved site-occupation profiles $n_{i\sigma}(t\leq 0)$ by means of a TDSDFT scheme in which the time-dependent xc potential is determined exactly at the ALSDA level (details see, Ref. [\onlinecite{gaoprb78Liwei}]). The performance of this method has been tested systematically against accurate t-DMRG simulation data for the repulsive Hubbard model.~\cite{gaoprb78Liwei} It is found that, the simple ALSDA for the time-dependent xc potential is surprisingly accurate in describing collective density and spin dynamics in strongly correlated 1D ultracold Fermi gases in a wide range of coupling strengths and spin polarizations. The performance of TDSDFT in describing the nonequilibrium behavior of strongly correlated lattice models has also been recently addressed in Ref. [\onlinecite{verdozzi_2007}].

In this work, we use this method to mainly discuss the nature of the interactions on the velocities of the density and spin evolution. The spin-charge dynamics after a local quench is discussed in Luttinger liquids (for $U>0$, gapless spin and charge excitations) and in Luther-Emery liquids (for $U<0$, gapless charge and gapful spin excitations). We consider at the same time the influence of polarization on the spin-charge dynamics. For attractive interactions, we limit our discussion on the weak-interaction case because for strong attractive interactions we found our SDFT code overestimates the amplitude of the bulk atomic density waves, which will greatly influence the TDSDFT results based on that.

Experimentally the strong local potential can be obtained by a blue- or red-detuned laser beam tightly focused
perpendicular to the 1D atomic wires, which generates locally repulsive or attractive potentials for the atoms in the wires, corresponding to
$W_\sigma>0$ or $W_\sigma<0$. In this paper, we are interested in the repulsive potential for the atoms.
The charge and spin densities can be observed by using {\it in situ} sequential absorption imaging, electron beams, or noise interference,~\cite{Shin} which, in principle, gives an unambiguous information on the spin-charge separation.

\section{Numerical results and discussion}
\label{sect:numerical_results}
In this section, we report on the results calculated by solving the time-dependent Kohn-Sham equations.
Mathematically the solution of the time-dependent Kohn-Sham
equations is an initial value problem. A given set of initial
orbitals calculated from the static Kohn-Sham equations is propagated forward in time. No self-consistent
iterations are required as in the static case.

For times $t \leq 0$, the system is in the presence of a strong local potential, which creates a strong local disturbance in ultracold gases and
makes the total density and spin-density distributions in the center of the system locally different (up to a few lattice sites).
We are interested in two kinds of quench dynamics. The first one is that, at time $t=0^+$, the local potential is quenched with the time-independent on-site interaction $U(t)=U$. The second is that, at time $t=0^+$, the local potential is switched off and at the same time the on-site interaction is quenched instantaneously with $U(t)=U\Theta(-t)$. After the quench, excitations are produced. We concern in this paper the subsequent real-time evolution of the spin-resolved densities after the quench, $n_{i\sigma}(t)=\langle\Psi(t)\vert \hat{n}_{i\sigma}\vert \Psi(t)\rangle$ with $\vert \Psi(t)\rangle$ the state of the system at time $t$. Charge density and spin density are defined accordingly as
$n_i(t)=n_{i\uparrow}(t)+n_{i\downarrow}(t)$ and $s_i(t)=[n_{i\uparrow}(t)-n_{i\downarrow}(t)]/2$.

If not mentioned otherwise, the numerical results presented below correspond to a system with $N_f=30$ atoms on $N_s=100$ sites, and with open (hard wall) boundary conditions imposed at the sites $i=0$ and $i=101$. The external potential is chosen to be spin dependent: $w_\uparrow=-1$ and $w_\downarrow=0$, used to form a local density and spin density occupations in the center of the system.

\subsection{$u>0$ and $p=0$}
\begin{figure}
\begin{center}
\includegraphics*[width=0.8\linewidth]{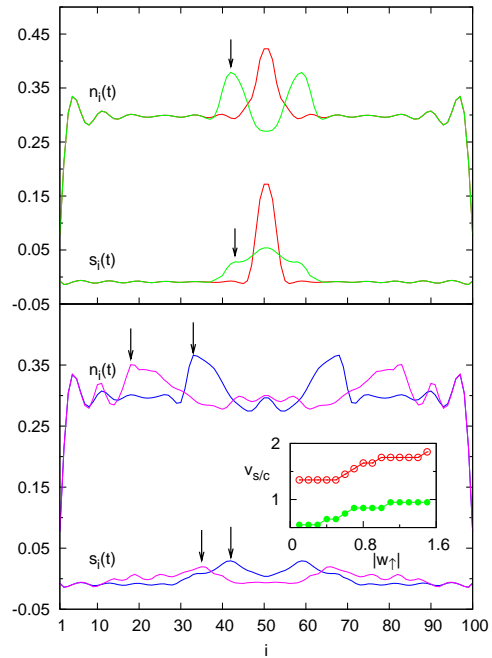}
\caption{(Color online) Charge $n_{i}(t)$ and spin $s_i(t)$ occupations as functions of lattice site $i$
and time $t$ for $N_s=100$, $N_\uparrow=N_\downarrow=15$, $w_\uparrow=-1$, $w_{\downarrow}=0$, $\alpha=2$, and repulsive interaction of $u=+2$.
Top panel: ground-state charge and spin occupations for times $t\leq 0$ (solid line) and  at time $t=5~\hbar/\gamma$ (dashed-dotted line). Bottom panel: same as in the top panels but at time $t=10~\hbar/\gamma$ (solid line) and  $t=20~\hbar/\gamma$ (dashed-dotted line). The charge and spin densities are plotted in the top and bottom of the panel, respectively.
The arrows in the plot indicate the positions where the wave packets propagate. In the inset, we show the velocities of the charge $v_c$ (open circles) and spin $v_s$ (solid circles) density wave packets as a function of the amplitude of the local potential $|w_\uparrow|$. Both velocities are increasing functions of $|w_\uparrow|$.
\label{fig:1}}
\end{center}
\end{figure}

In Fig. \ref{fig:1}, we show results for a spin-unpolarized system ($N_\uparrow=N_\downarrow=15$) with repulsive interaction of $u=2$.
At $t \le 0$, a dominant local charge- and spin-density profiles in the center of the system are generated by the strong local potential.
After the quench of the local potential, the charge and spin densities evolve and split into two counterpropagating density wave packets.
The propagation in time is in fact due to the nonequilibrium initial condition. The charge density evolves with a quicker velocity than the spin, which is in agreement with the general picture of spin-charge separation.~\cite{giamarchi_book} A qualitative analysis based on the continuity equation for the momentum density can also well explain the phenomena of spin-charge separation.~\cite{gaoprl102}

We notice a common feature in almost all the figures in this paper, that is, the spin and charge
densities have an asymmetric forward-leaning shape. This is caused by a nonlinear effect, i.e., the different local velocities in the center and at the edges. Since the local velocity is proportional to the density, the higher density in the center gains larger velocity than that at the edges, which qualitatively explains why
the asymmetric forward-leaning shape happens during the density propagation. For perturbations with small amplitude, the charge velocity is studied in details by t-DMRG and compared to
the Bethe-ansatz results with good agreement.~\cite{kollath} For the strong local potential studied here, the spin and charge velocities, determined from the propagation of the maximum of the charge and spin wave packets away from the center, vary with time. We thus calculate and compare the velocities determined at the fixed time $t=10 \hbar/\gamma$. In the inset of Fig. \ref{fig:1}, we show the spin and charge velocities as a function of the amplitude of the local potential $|w_\uparrow|$. We find both velocities are increasing functions of $|w_\uparrow|$.
For the charge background density ($\sim 0.3$) in Fig. \ref{fig:1}, the charge and spin velocities by the Bethe-ansatz method are $v_c=1.15$ and $v_s=0.75$. In the limit of $w_\uparrow \rightarrow 0$, but $w_\downarrow \equiv 0$, our results give $v_c=1.3$ and $v_s=0.65$. The differences are possibly caused by the simultaneous local perturbations in the charge and spin densities used here, which break the spin-charge scenario and couple the spin and charge modes, similar to the effects caused by the finite spin polarization (see Secs. III-C and III-D).

\begin{figure}
\begin{center}
\tabcolsep=0 cm
\includegraphics[width=1.0\linewidth]{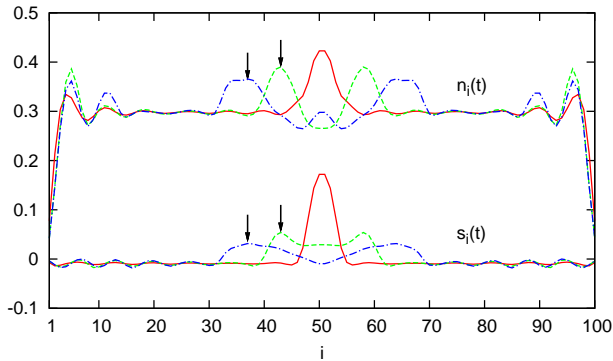}
\caption{(Color online)  Charge $n_i(t)$ and spin $s_i(t)$ occupations as functions of lattice site $i$ and time $t$ with quenches for
the local potential and on-site interaction, i.e., $V_{i\sigma}(t)=V^{\rm ext}_{i\sigma} \Theta(-t)$ and $U(t)=U\Theta(-t)$. The other parameters are the same as that in Fig. \ref{fig:1}. The static density (solid line) is shown together with two time shots for $t=5~\hbar/\gamma$ (dash line) and  $t=10~\hbar/\gamma$ (dashed-dotted line).
\label{fig:2}}
\end{center}
\end{figure}
\begin{figure}
\begin{center}
\includegraphics*[width=1.0\linewidth]{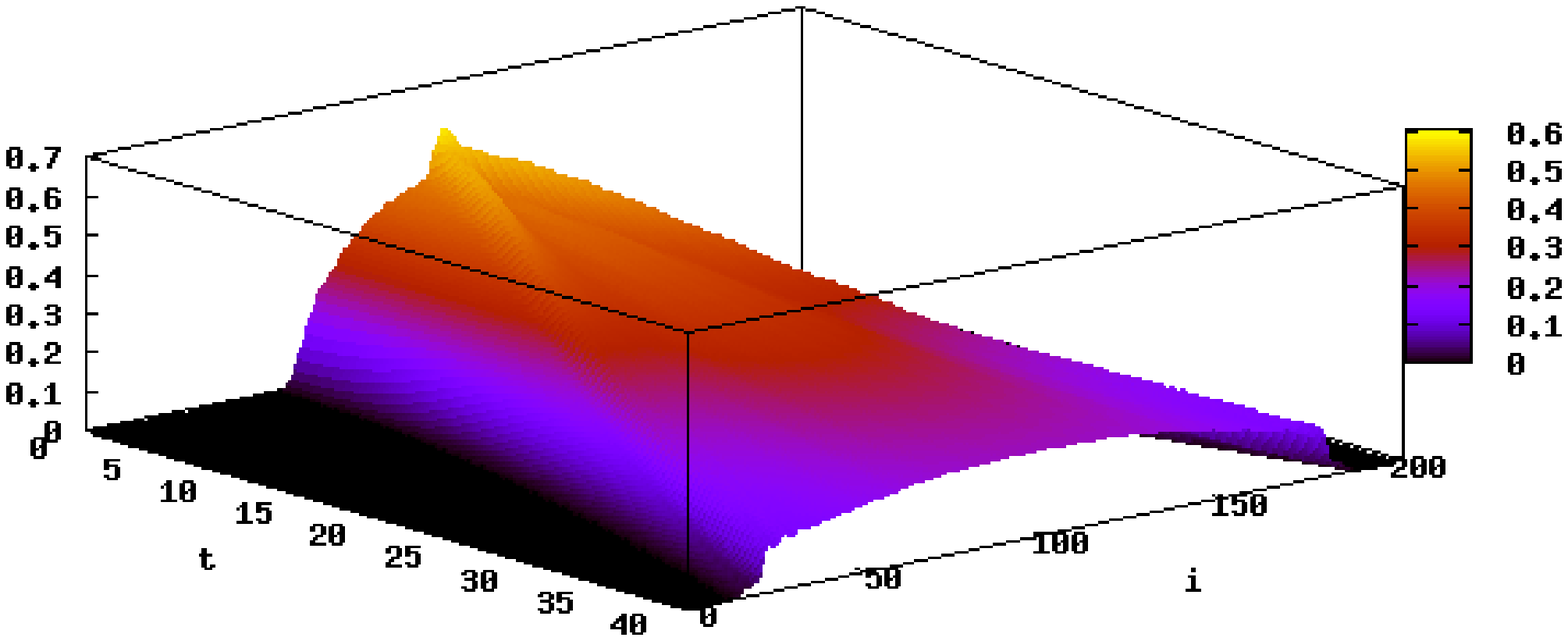}\\
\includegraphics*[width=1.0\linewidth]{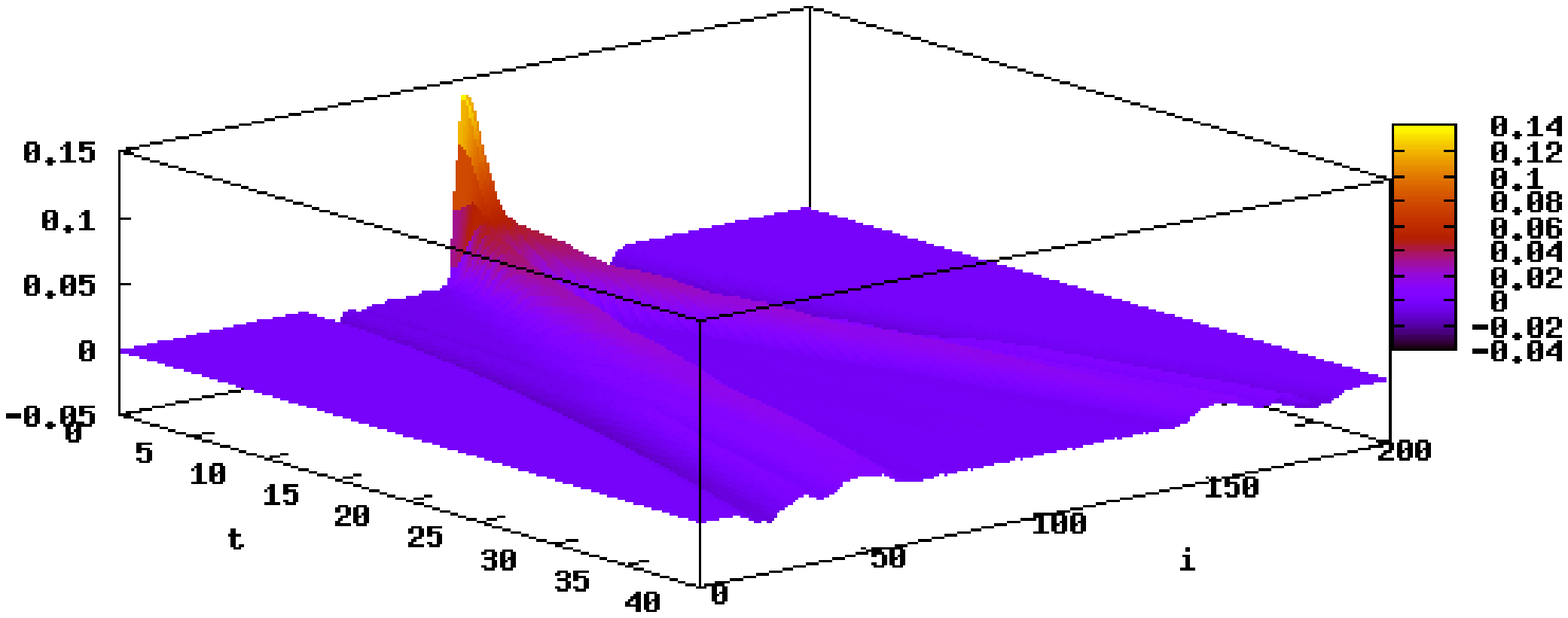}
\caption{(Color online) 3D plots for the Charge density $n_{i}(t)$ (Top panel) and spin density $s_i(t)$ (Bottom panel) as functions of lattice site $i$
and time $t$ (in units of $\hbar/\gamma$) for a harmonically trapped system with $N_s=200$, $N_\uparrow=N_\downarrow=15$, $V_2/\gamma=5\times 10^{-4}$, $w_\uparrow=-1$, $w_{\downarrow}=0$, $\alpha=2$, and repulsive interaction of $u=2$.
\label{fig:3}}
\end{center}
\end{figure}

In Fig. \ref{fig:2}, we study the local potential quench together with an on-site interaction quench, i.e.,
$V_{i\sigma}(t)=V^{\rm ext}_{i\sigma} \Theta(-t)$ and $U(t)=U\Theta(-t)$. We find that, the spin- and charge-density wave packets split
and counterpropagate as usual but the phenomena of the spin-charge separation completely disappears. That is, the spin and charge densities evolve with the same velocity. From the
Luttinger-liquid theory based on the bosonization method~\cite{Coll} or from the Bethe-ansatz solution,~\cite{Schulz} one can derive that the spin velocity $v_s$ and the charge velocity $v_c$ satisfy $v_c=v_s=v_F$ in the noninteracting limit, with $v_F=2\gamma\sin(\pi n/2)$ the Fermi velocity. The interaction between the different species is one of the important ingredients for the spin-charge separation, which explains the suppression of the spin-charge separation after the interaction quench. Making use of the techniques from the cold atomic gases, two different ways of quenching, used in Figs. \ref{fig:1} and  \ref{fig:2}, respectively, can give a clear signal that
different collective spin and charge dynamics happens when starting from the same initial strong local perturbation. We would like to mention that Kollath proposed to repeat the dynamics in Fig. \ref{fig:1} in higher dimensions where no separation of spin and charge should be seen.~\cite{Kollath2} We notice that in Fig. \ref{fig:2} already at short time, some density waves coming from the sharp edges begin to influence the charge- and spin-density wave packets from the center. At larger time, they will mix with the original packets.
\begin{figure}
\begin{center}
\includegraphics*[width=0.8\linewidth]{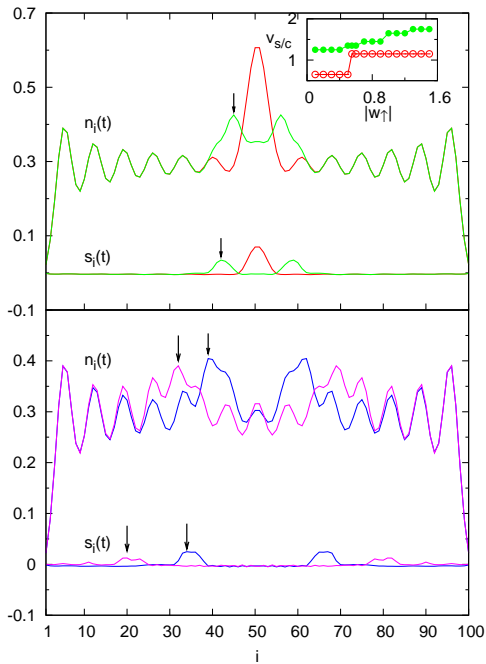}
\caption{(Color online) Charge $n_{i}(t)$ and spin $s_i(t)$ occupations as functions of lattice site $i$
and time $t$ for $N_s=100$, $N_\uparrow=N_\downarrow=15$, $w_\uparrow=-1$, $w_{\downarrow}=0$, $\alpha=2$, and attractive interaction of $u=-1$.
Top panel: ground-state charge and spin occupations for times $t\leq 0$ (solid line) and  at time $t=5~\hbar/\gamma$ (dashed-dotted line). Bottom panel: same as in the top panels but at time $t=10~\hbar/\gamma$ (solid line) and  $t=20~\hbar/\gamma$ (dashed-dotted line). The inset shows the velocities of the charge $v_c$ (open circles) and spin $v_s$ (solid circles) density wave packets as a function of the amplitude of the local potential $|w_\uparrow|$.
\label{fig:4}}
\end{center}
\end{figure}
\begin{figure}
\begin{center}
\includegraphics*[width=1.0\linewidth]{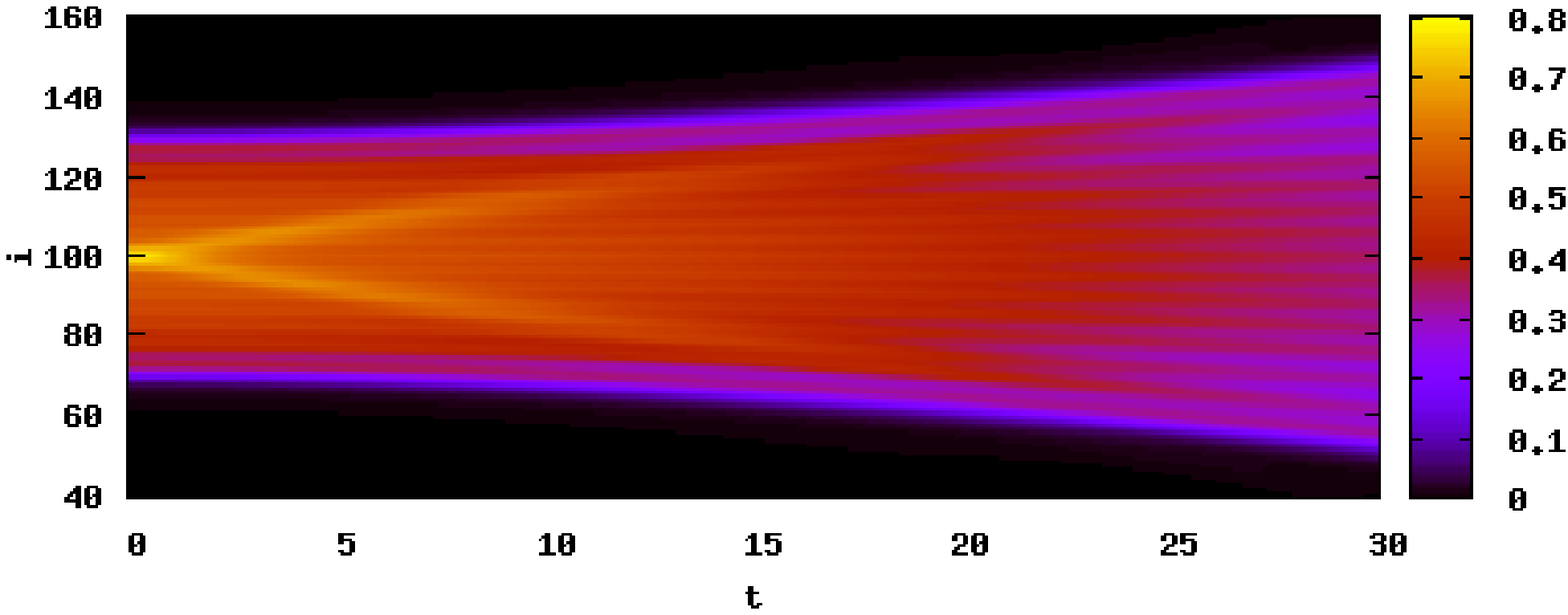}\\
\includegraphics*[width=1.0\linewidth]{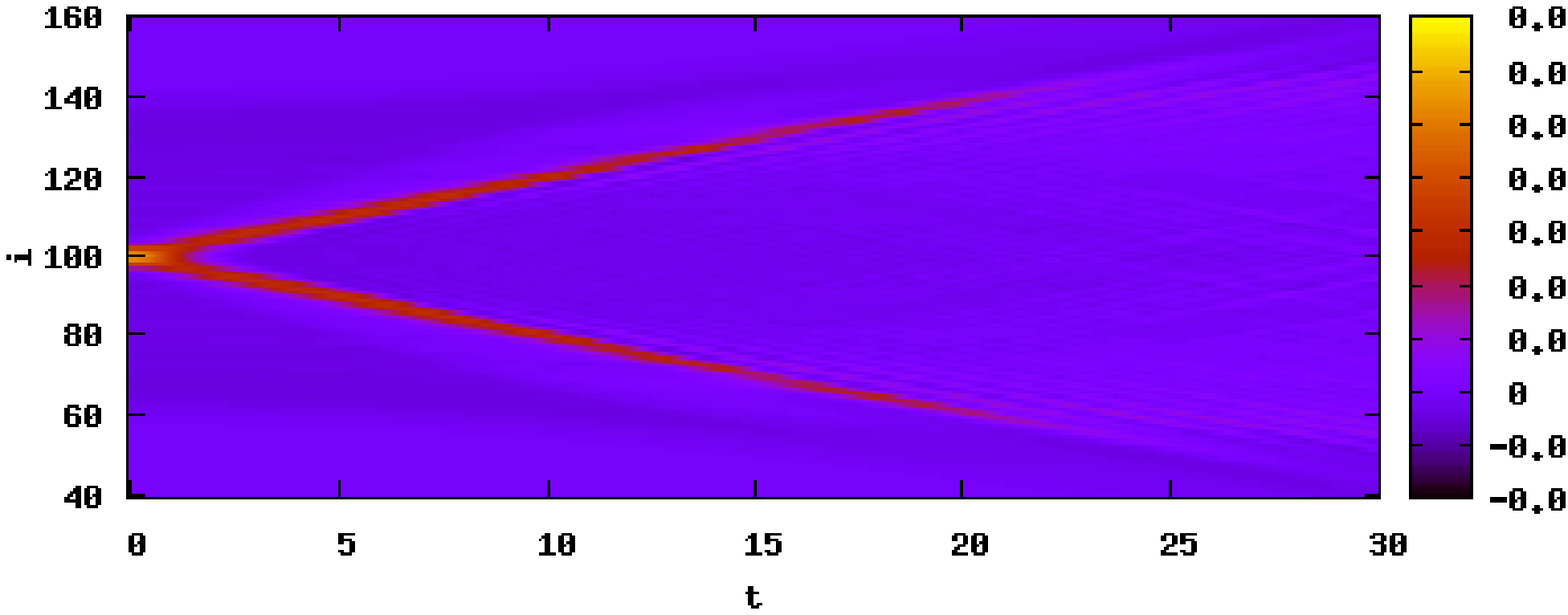}
\caption{(Color online) Contour plots for the Charge density $n_{i}(t)$ (Top panel) and spin density $s_i(t)$ (Bottom panel) as functions of lattice site $i$
and time $t$ (in units of $\hbar/\gamma$) for a harmonically trapped system with $N_s=200$, $N_\uparrow=N_\downarrow=15$, $V_2/\gamma=5\times 10^{-4}$, $w_\uparrow=-1$, $w_{\downarrow}=0$, $\alpha=2$, and attractive interaction of $u=-1$.
\label{fig:5}}
\end{center}
\end{figure}

In practice, an additional trapping potential is unavoidable in the present experimental set-ups. We thus present our simulations for the system in the presence of an additional weak superimposed harmonic trapping potential, namely, $V^{\rm ext}_{i\sigma}$ in Eq. (\ref{eq:ext_pot}) is changed into,
\begin{eqnarray}\label{eq:ext_pot1}
V^{\rm ext}_{i\sigma}=W_\sigma e^{-\frac{[i-(N_s+1)/2]^2}{2\alpha^2}}+ V_2 \left(i-\frac{N_s+1}{2}\right)^2.
\end{eqnarray}
Here we take $V_2/\gamma=5\times 10^{-4}$. The three-dimensional (3D) plots of the time evolution of the spin- and charge-density wave packets are shown in Fig. \ref{fig:3}. From the figure, we observe that,
in the presence of the harmonic potential the charge and spin wave packets are highly inhomogeneous, but the counter-propagation and the separation of the charge- and spin-density wave packets are still visible in the background of the inverted parabola.

\subsection{$u<0$ and $p=0$}
In one-dimensional Hubbard model, away from half filling, the spin and charge velocities of the low-energy collective excitations
satisfy,~\cite{Coll,Schulz}
\[
v_{s,c}=v_F\sqrt{1\mp \frac{U}{\pi v_F}}\,.
\]
This gives a qualitative explanation that for the positive-$U$ Hubbard model, the charge velocity is larger than the spin velocity, while
for the negative-$U$ Hubbard model, the charge velocity is smaller than the spin velocity. In Fig. \ref{fig:4}, the quench dynamics for the attractive
Hubbard model, which belongs to the Luther-Emery universality class, illustrates that spin-wave packets evolve with a faster speed than the charge branches.
In the inset of Fig. \ref{fig:4}, we show the spin and charge velocities evaluated at $t=10 \hbar/\gamma$ as a function of the amplitude of the local potential $|w_\uparrow|$.
We notice that an abrupt change appears in the charge velocity at $|w_\uparrow|\approx 0.55$.
For attractive interactions, Luther-Emery paring induces a prominent density wave characterized by the dip-hump structure. While the charge velocity is determined from the propagation of the maximum of the charge wave packets located at one of the humps of the density wave. The increase in the amplitude of the local potential makes the maximum of the charge wave packets move from the lattice site $i=44$ to $i=39$, which explains the discontinuity of the charge velocity for attractive interactions at $|w_\uparrow|\approx 0.55$.
However, this discontinuous change has artifacts because the way of extracting $v_{c,s}$ used here is not an optimum one.
In Fig. \ref{fig:5}, we present the contour plots of the time evolution of the density and spin packets for the system in the presence of a harmonic trapping potential with $V_2/\gamma=5\times 10^{-4}$. The different evolution velocities for the charge- and spin-density wave packets are clearly visible.
\begin{figure}
\begin{center}
\tabcolsep=0 cm
\includegraphics[width=1.0\linewidth]{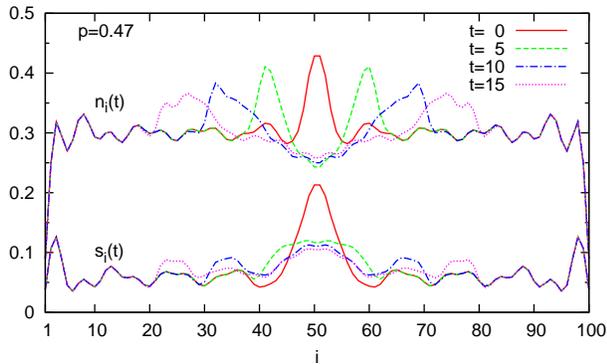}
\caption{(Color online) The ground-state charge and spin occupations as functions of lattice site $i$ and time $t$ for the system of repulsive interaction of $u=2$ in the polarized case of $P=0.47$ ($N_\uparrow=22, N_\downarrow=8$). Besides the ground-state density and spin density (solid line), three time shots are shown with $t=5~\hbar/\gamma$ (dash line,) $t=10~\hbar/\gamma$ (dashed-dotted line), and  $t=15~\hbar/\gamma$ (dotted line).
\label{fig:6}}
\end{center}
\end{figure}
\begin{figure}
\begin{center}
\tabcolsep=0 cm
\includegraphics[width=1.0\linewidth]{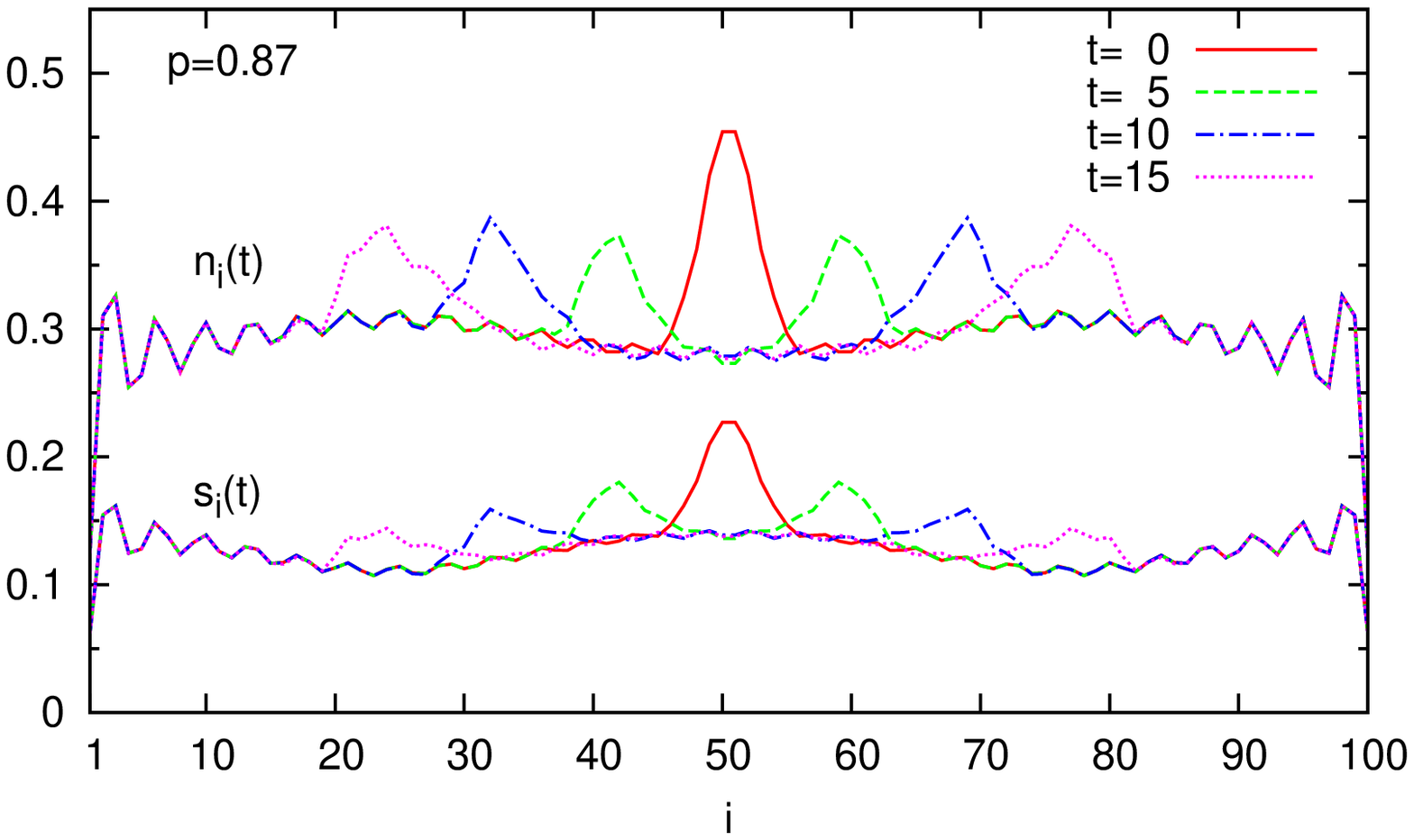}
\caption{(Color online) Same as Fig. 6 but for the polarized system of $P=0.87$ ($N_\uparrow=28, N_\downarrow=2$).
\label{fig:7}}
\end{center}
\end{figure}
\begin{figure}
\begin{center}
\tabcolsep=0 cm
\includegraphics[width=1.0\linewidth]{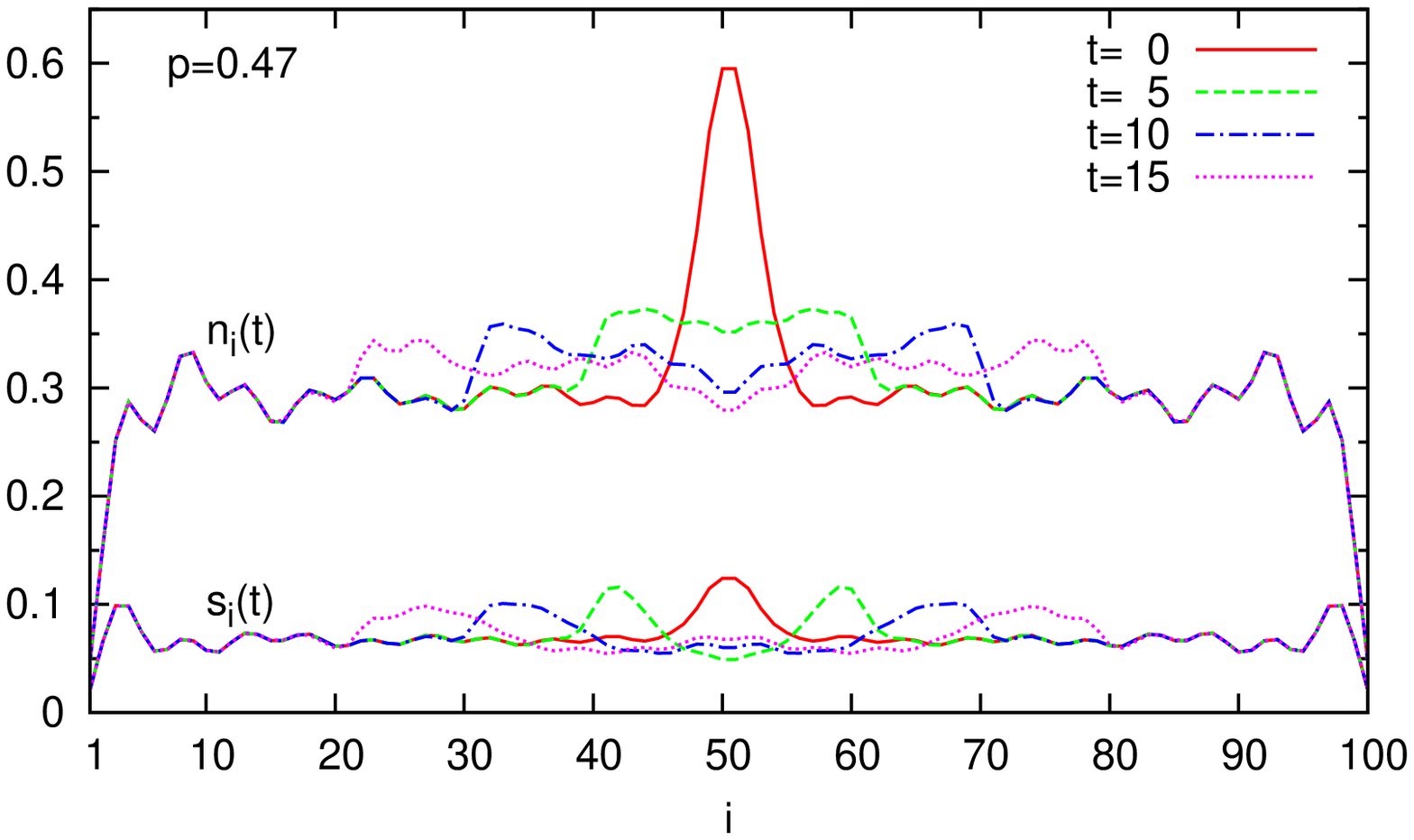}
\caption{(Color online) The ground-state charge and spin occupations as functions of lattice site $i$ and time $t$ for the system of attractive interaction of $u=-1$ in the polarized case of $p=0.47$ ($N_\uparrow=22, N_\downarrow=8$). Besides the ground-state density and spin density (solid line), three time shots are shown with $t=5~\hbar/\gamma$ (dash line,) $t=10~\hbar/\gamma$ (dashed-dotted line), and  $t=15~\hbar/\gamma$ (dotted line).
\label{fig:8}}
\end{center}
\end{figure}
\begin{figure}
\begin{center}
\tabcolsep=0 cm
\includegraphics[width=1.0\linewidth]{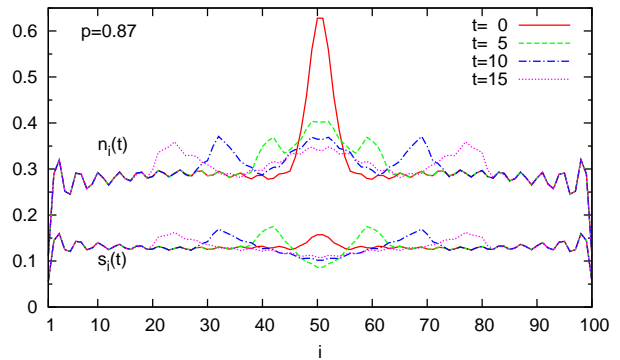}
\caption{(Color online) Same as Fig. 8, but for the polarized system of $p=0.87$ ($N_\uparrow=28, N_\downarrow=2$).
\label{fig:9}}
\end{center}
\end{figure}

\subsection{$u>0$ and $p\ne 0$}
The spin-charge separation in a spin-polarized one-dimensional system is quite different from the fully polarized one.
The spin-charge-coupled dynamics in a polarized system formulated with the first-quantized path-integral formalism and bosonization techniques
provides us a new non-Tomanaga-Luttinger-liquid universality class.~\cite{Akhanjee}
For the Luther-Emery liquid of unpolarized attractive Fermi gases, the spin and charge degrees of freedom are decoupled.
In contrast, in the system with finite spin imbalance, spin-charge mixing is found based on an effective-field theory for the long-wavelength and low-energy
properties.~\cite{Erhai} In Figs. \ref{fig:6} and \ref{fig:7}, the quench dynamics for spin- and charge-density waves is shown for the system of repulsive interaction ($u=2$) with polarization of $p=0.47$ and $0.87$, respectively. For $p\geq 0.47$, there is only small difference between spin and charge velocities. In the case of a large polarization, the same propagating velocities for spin and charge are obtained.

\subsection{$u<0$ and $p\ne 0$}
The quench dynamics for spin and charge density waves of the attractive case for $u=-1$ is shown in Figs. \ref{fig:8} and \ref{fig:9}. We find with the increasing of the polarization, the spin-charge separation is strongly suppressed due to the interplay between charge and spin degrees of freedom. Theoretically, for the partially polarized system, the spin and charge modes are coupled. In this case, there is no strict spin-charge separation scenario, namely, the spin-charge separation breaks down. Numerically, we observe that, at small polarization the spin and charge wave packets still evolve at different velocities although they are coupled and influence each other. At large polarization, the spin-charge separation disappears and evolves at the same velocities for both the repulsive and the attractive systems we studied.

\section{Conclusions}
\label{sect:conclusions}
In summary, we have calculated the non-equilibrium dynamic evolution of a one-dimensional system of two-component fermionic atoms after a strong local quench with or without interaction quench by using a time-dependent density-functional theory with a suitable Bethe-ansatz based adiabatic local spin-density approximation. A test of the performance of TDSDFT is provided for the unpolarized systems with attractive or repulsive interactions in the presence of a harmonic trapping potential. Under the same local perturbation, the charge velocity is larger than the spin velocity for the system of repulsive interaction and vice versa for the attractive case, which is compatible with the low-energy collective dynamics from the Bethe-ansatz solution or the bosonization techniques. We found the spin-charge separation is strongly suppressed when the interaction quench is forced together with the local potential quench. Spin-charge mixing is found for the system of polarization signaling by the disappearance of the spin-charge separation. Numerically we observe that the spin-charge separation disappears for large polarizations in both the repulsive and the attractive Hubbard model we studied.

\acknowledgments
This work was supported by NSF of China under Grants No. 10974181 and No. 10704066, Qianjiang River Fellow Fund 2008R10029, Program for Innovative Research Team in Zhejiang Normal University, and partly by the Project of Knowledge Innovation Program (PKIP) of Chinese Academy of Sciences under Grant No. KJCX2.YW.W10.

\end{document}